\documentclass[aps,superscriptaddress,amsmath,amssymb,floatfix,twocolumn,showpacs,amsfonts,longbibliography,prl]{revtex4-1}
\usepackage{times}
\usepackage[varg]{txfonts}
\usepackage{textcomp}
\usepackage{graphicx}
\usepackage{subfigure}
\usepackage{tabu}
\usepackage{color}
\usepackage[colorlinks=true,citecolor=blue,urlcolor=blue,linkcolor=blue,hyperindex]{hyperref}
\usepackage{braket}
\usepackage{float}
\usepackage{overpic}
\usepackage{amssymb}
\usepackage{amsmath}
\usepackage{gensymb}
\usepackage{mathrsfs}
\usepackage{tikz}
\usepackage{bm}

\allowdisplaybreaks

\begin{document}

\title{Deconfined quantum phase transition on the Kagome lattice: Distinct velocities of spinon and string excitations}
\author{Dong-Xu Liu}
\affiliation{Department of Physics, Chongqing University, Chongqing 401331, China}
\author{Zijian Xiong}
\email{Corresponding author: xiongzj@cqu.edu.cn}
\affiliation{Department of Physics, Chongqing University, Chongqing 401331, China}
\affiliation{Department of Applied Physics, University of Tokyo, Tokyo 113-8656, Japan}
\affiliation{College of Physics and Electronic Engineering, Chongqing Normal University, Chongqing 401331, China}
\author{Yining Xu}
\affiliation{College of Physics and Electronic Engineering, Chongqing Normal University, Chongqing 401331, China}
\author{Xue-Feng Zhang}
\email{Corresponding author: zhangxf@cqu.edu.cn}
\affiliation{Department of Physics, Chongqing University, Chongqing 401331, China}

\begin{abstract}
Deconfined quantum phase transition (DQPT) provides an extraordinary possibility of the quantum phase transition beyond the Ginzburg-Landau paradigm, which is interwoven with numerous exotic phenomena of the strongly correlated quantum many-body system, e.g. fractional excitation, emergent symmetries, and gauge field. However, various candidates of DQPT have been demonstrated to be weakly first-order, and the conformal field theory (CFT) has to be altered into a non-unitary one. 
Here we numerically found two linear dispersions with different velocities in one of the few survivors of DQPT --- the extended hard-core Bose-Hubbard model on the Kagome lattice. Such counterintuitive results directly lead to the negation of possible emergent Lorentz symmetry, and the breakdown of conventional theory of DQPT. Furthermore, the snapshots of boson configuration hint that these two velocities may correspond to the dynamics of the fractional excitations and quantum strings, respectively. Our work will inspire the revisit of the theory of DQPT and benefit the field of quantum materials and quantum simulations.
\end{abstract}
\maketitle
\date{\today}
\underline{Introduction}---The quantum phases breaking different symmetries usually undergo a discontinuous transition between them, but DQPT provides a prototypical counter-example, e.g. the superfluid (SF) phase can undergo a continuous transition to the valence bond solid (VBS) \cite{dqcp,dqcplong,dqcp_sandvik,dqcp_adam,dqcp_WC,dqcp_ZYM,dqcp_assaad,dqcp_zhang}. The low-energy physics of DQPT can be interpreted with two flavor spinons coupled to the U(1) gauge field \cite{dqcp,dqcplong}. The duality web proposes  the deconfined quantum critical point (DQCP) is either conformal invariant or flowing into the non-unitary CFT \cite{ChongWang2017prx}. Recent researches show that most DQCP candidates belong to the non-unitary CFT \cite{do_yangcheng,do_assaad,uCFT_WC,nCFT}. Importantly, the proximate SU(2) DQPT is recently found in the real material SrCu$_{2}$(BO$_{3}$)$_{2}$ \cite{dqcp_yu}.

The DQPT on the Kagome lattice belongs to the easy-plane type \cite{kagome_Senthil,kagome_wessel,kagome_zhang,dqcp_zhang,dqcp_frozen}. According to the field theory  \cite{ChongWang2017prx} and RG analysis \cite{dqcp_frozen}, the effective low-energy model is self-dual and supports continuous transition, so the conformal invariant is expected. Numerically, the DQPT on the Kagome lattice also supports the CFT, such as the critical exponents $\eta_{\mathrm{VBS}}$ and $\eta_{\mathrm{SF}}$ are very close, and the scaling dimension of spinons is two. It is straightforward to believe the dynamics of the system also support the conformal invariant.

\begin{figure}[t]
	\centering
	\includegraphics[width=0.48\textwidth]{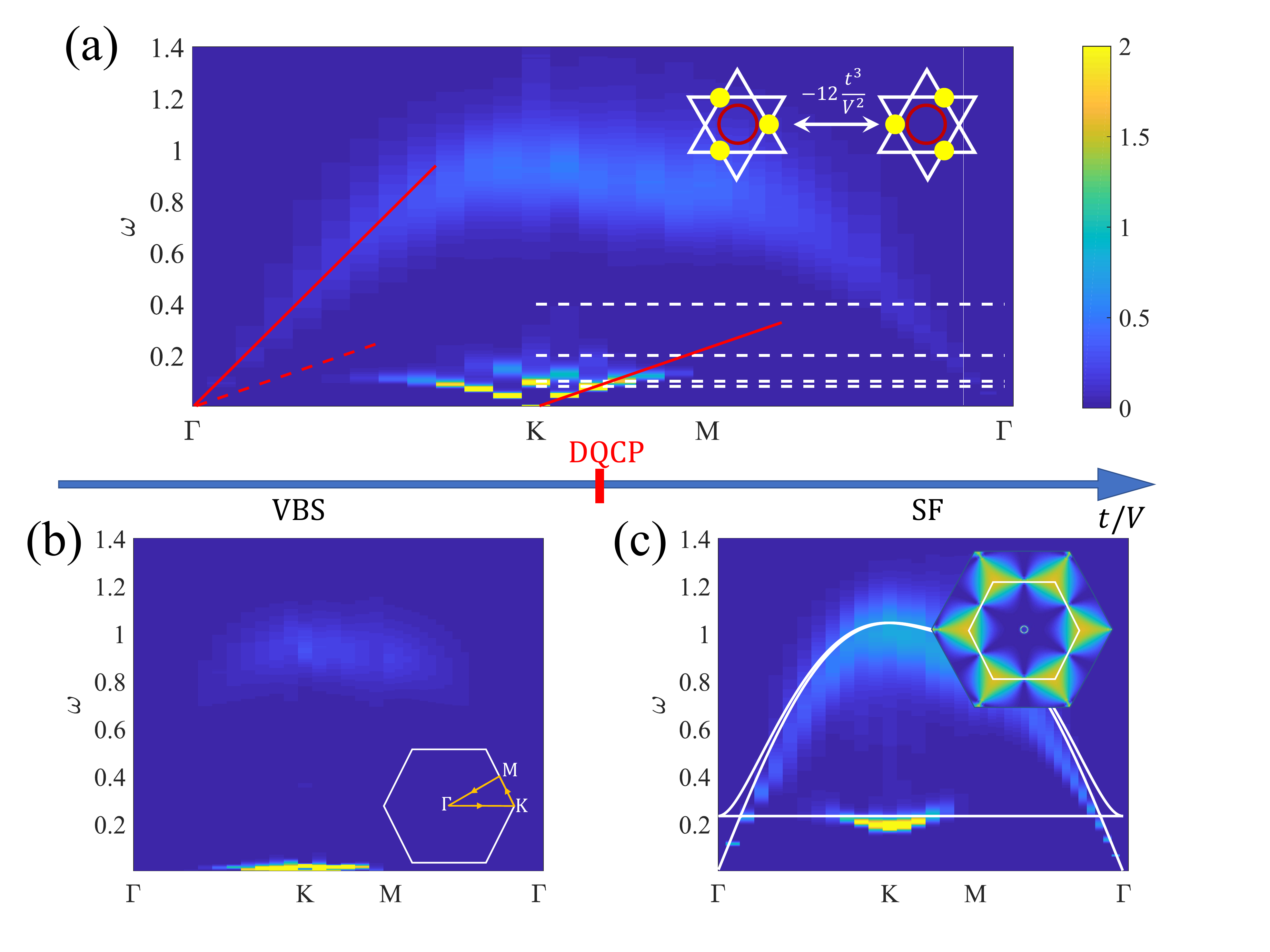}
	\caption{The dynamic spectra of (a) DQCP at $t/V=0.1302$, (b) VBS phase at $t/V=0.08$, and (c) SF phase at $t/V=0.20$ calculated with QMC simulation ($L=36$). The inset of (a) presents ring exchange, the red solid lines show the fitting speeds at $\Gamma$ and K point (red dash line for comparison), and the white dash lines highlight the parameter regions in Fig.\ref{fig2}. The scanning path $\Gamma$-K-M-$\Gamma$ is depicted in the inset of (b). The solid white lines in (c) are energy dispersion calculated by LSWT. The inset of (c) is DSF calculated by LSWT with constant energy cut close to the flat band.}\label{fig1}
\end{figure}

In this paper, we study the dynamics of the DQPT between VBS and SF phases on the Kagome lattice with the large-scale quantum Monte Carlo (QMC) method. As shown in Fig. \ref{fig1} (a), two types of linear dispersions are observed at $\Gamma$ and K points, respectively. The large deviation between their speeds is verified after careful finite-size scaling analysis, indicating the non-existence of Lorentz symmetry at the DQCP which conflicts with the prediction of CFT.

\underline{Model}---We consider the extended hard-core Bose-Hubbard Hamiltonian \cite{kagome_Senthil,kagome_wessel,kagome_zhang,dqcp_zhang} written as 
\begin{equation}\label{xxzmo}
H=-t\sum_{\langle i,j\rangle}(b^{\dagger}_{i}b_{j}+h.c.)+V\sum_{\langle i,j\rangle}n_{i}n_{j},
\end{equation}
where $t\ge0$ and $V\ge0$ denote the hopping and repulsive interaction between nearest-neighbor sites, and $b^{\dagger}_{i}$ ($b_i$) is the creation (annihilation) operator of the hard-core boson which could be used for describing the Rydberg-dressed atom in the optical lattice \cite{Rydberg-dressed}. After implementing the mapping $b^{\dagger}_i\rightarrow S^+_i$, $b_i\rightarrow S^-_i$ and $n_i\rightarrow S_i^z+1/2$, the Hamiltonian is equivalent to the spin-half XXZ model which is related to the quantum magnetism \cite{NIS_2012}. Because the DQPT happens at an average density equal to $1/3$, the numerical simulation is performed in the canonical ensemble \cite{dqcp_zhang}. Here, we choose $V$ as the energy unit.

Minimization of the ground state energy imposes a strong local constraint --- each triangle of the Kagome lattice can be occupied by only one particle (e.g. ${\tikz{\fill (0,0) circle (0.5mm);\draw (0,0) -- (0.1,0.1732) -- (0.2,0) -- (0,0);}}$ or ${\tikz{\fill (0.1,0) circle (0.5mm);\draw (0,0.1732) -- (0.2,0.1732) -- (0.1,0) -- (0,0.1732);}}$ named triangle rule). Similar to the spin ice \cite{ice_Balents}, the ground state is disordered with macroscopic degeneracy \cite{Pauling1935}. However, in the strong coupling region $t/V \ll 1$, a third-order perturbative interaction $H_{\textrm{ring}} = -\frac{12t^{3}}{V^{2}}\sum (|{\tikz[scale=0.8]{\fill (0,-0.1) circle (0.5mm);\fill (0,0.2464) circle (0.5mm);\fill (0.3,0.0732) circle (0.5mm);\draw (0,-0.1) -- (-0.1,0.0732) -- (0,0.2464) -- (0.2,0.2464) -- (0.3,0.0732) -- (0.2,-0.1) -- (0,-0.1);}}\rangle \langle \tikz[scale=0.8]{\fill (-0.1,0.0732) circle (0.5mm);\fill (0.2,0.2464) circle (0.5mm);\fill (0.2,-0.1) circle (0.5mm);\draw (0,-0.1) -- (-0.1,0.0732) -- (0,0.2464) -- (0.2,0.2464) -- (0.3,0.0732) -- (0.2,-0.1) -- (0,-0.1);} |+h.c.)$ can exchange the configuration without breaking the triangle rule (inset of Fig.\ref{fig1}(a)) and lift the degeneracy so that the system enters the VBS phase. Although some spinons (e.g. ${\tikz{\fill (0,0) circle (0.5mm);\fill (0.1,0.1732) circle (0.5mm) ;\draw (0,0) -- (0.1,0.1732) -- (0.2,0) -- (0,0);}}$ or ${\tikz{\draw (0,0) -- (0.1,0.1732) -- (0.2,0) -- (0,0);}}$) can still be excited due to quantum fluctuations, the large energy gap $V$ makes them confined.

In the weak coupling region $t \gg V$, the large hopping process makes the bosons break the U(1) symmetry so that the system enters the SF phase. As shown in Fig.1(c) with the help of linear spin wave theory (LSWT), along the selected path $\Gamma$-K-M-$\Gamma$, the lowest gapless branch at $\Gamma$ point with the linear dispersion corresponds to the Goldstone mode, and the flat-band branch results from the lattice geometry \cite{flatband}.

At the critical point, the interplay between the spinons and the emergent dynamical U(1) gauge field leads to the deconfined criticality. Several exotic phenomena can be found, such as the drift of the superfluid density, large anomalous critical exponent, emergent symmetries, and so on \cite{dqcp_zhang}. It is a common belief that this DQPT can be described by the easy plane NCCP$^1$ theory \cite{dqcp,dqcplong}.

{\underline{Spectra of phases}}---The numerical method adopted is the stochastic cluster series expansion with parallel tempering \cite{scse_pt} which can greatly overcome the non-ergodic problem. We choose the periodic boundary condition (PBC) with the largest system size reaching $N=36\times36\times3=3888$ sites. Meanwhile, in order to suppress the influence of the thermal fluctuation, the temperature is set to be $T=1/\beta=\frac{6V}{100L}$ which is even lower than our previous work \cite{dqcp_zhang}. Here, we focus on the dynamical structure factor (DSF)  $S^{zz}(\textbf{k},\omega)=\frac{1}{2\pi L^2}\sum_{ij}\int_{-\infty}^{+\infty}\mbox{d}t\,e^{i\mathbf{k}\cdot(\mathbf{r}_{i}-\mathbf{r}_{j})-i\omega t}\langle S_{i}^{z}(0)S_{j}^{z}(t)\rangle$, which can be extracted from the imaginary time correlation function $S^{zz}(\textbf{r},\tau)=\langle S^z(\textbf{0},0) S^z(\textbf{r},\tau)\rangle$ by implementing the stochastic analytic continuation (SAC) method \cite{sac_review,sac_shao,sac_dqcp1,sac_string,ice_sac}. The QMC samples are more than five million, so high-quality spectra can be obtained.

In the VBS phase, two branches can be observed in Fig.\ref{fig1} (b). The lower branch stays on a very low energy scale and is nearly flat with a tiny gap (see appendix). The flat band is usually related to the lattice geometry, and it reflects the localization of the particles caused by the effective ring exchange interaction \cite{flatband}. Similar to the checkerboard lattice \cite{ice_sac}, it corresponds to the excitation from triplet ground state $\frac1{\sqrt{2}}(|{\tikz[scale=0.8]{\fill (0,-0.1) circle (0.5mm);\fill (0,0.2464) circle (0.5mm);\fill (0.3,0.0732) circle (0.5mm);\draw (0,-0.1) -- (-0.1,0.0732) -- (0,0.2464) -- (0.2,0.2464) -- (0.3,0.0732) -- (0.2,-0.1) -- (0,-0.1);}}\rangle +| \tikz[scale=0.8]{\fill (-0.1,0.0732) circle (0.5mm);\fill (0.2,0.2464) circle (0.5mm);\fill (0.2,-0.1) circle (0.5mm);\draw (0,-0.1) -- (-0.1,0.0732) -- (0,0.2464) -- (0.2,0.2464) -- (0.3,0.0732) -- (0.2,-0.1) -- (0,-0.1);} \rangle)$ to the singlet state $\frac1{\sqrt{2}}(|{\tikz[scale=0.8]{\fill (0,-0.1) circle (0.5mm);\fill (0,0.2464) circle (0.5mm);\fill (0.3,0.0732) circle (0.5mm);\draw (0,-0.1) -- (-0.1,0.0732) -- (0,0.2464) -- (0.2,0.2464) -- (0.3,0.0732) -- (0.2,-0.1) -- (0,-0.1);}}\rangle -| \tikz[scale=0.8]{\fill (-0.1,0.0732) circle (0.5mm);\fill (0.2,0.2464) circle (0.5mm);\fill (0.2,-0.1) circle (0.5mm);\draw (0,-0.1) -- (-0.1,0.0732) -- (0,0.2464) -- (0.2,0.2464) -- (0.3,0.0732) -- (0.2,-0.1) -- (0,-0.1);} \rangle)$, so its energy scale is $\sim\frac{12t^{3}}{V^{2}}$. Meanwhile, one can see the disappearance of the flat band along the high symmetry line $\Gamma$-M. Such fragmentation implies the existence of ``selection rule" \cite{ice_sac,string_xiong}. On the other hand, the higher branch has a large energy gap $\sim V$, and it is relevant to the spinon separation due to the local quantum fluctuation.

In the SF phase, the numerical spectrum in Fig.\ref{fig1}(c) also shows two branches. The energy scale of the flat band branch increases a lot, while the fragmentation remains. From the LSWT calculation of DSF at the flat band (inset of Fig.\ref{fig1}(c)), we find no intensity along the path M-$\Gamma$ which is consistent with the numerical results. We think this fragmentation should also result from some ``selection rule" due to the lattice symmetries \cite{ice_sac,string_xiong}. On the other hand, the higher branch is the gapless Goldstone mode.

{\underline{Spectra of DQCP}}---The VBS and SF phases break different symmetries but can still undergo a continuous phase transition at DQCP between them.  As shown in Fig.\ref{fig1} (a) at DQCP, the higher branch (named $\Gamma$-branch) is gapless at $\Gamma$ point and has a similar shape as the SF phase but with lower intensity. Deformed from the Goldstone mode in the SF phase, the $\Gamma$-branch becomes more continuous in a high energy scale, and it indicates the emergence of the fractional charges at DQCP. Notice that, here we do not choose the logarithmic scale in the $\omega$ axis, so the continuum of DSF may appear less obvious.
\begin{figure}[t]
	\centering
	\includegraphics[width=0.48\textwidth]{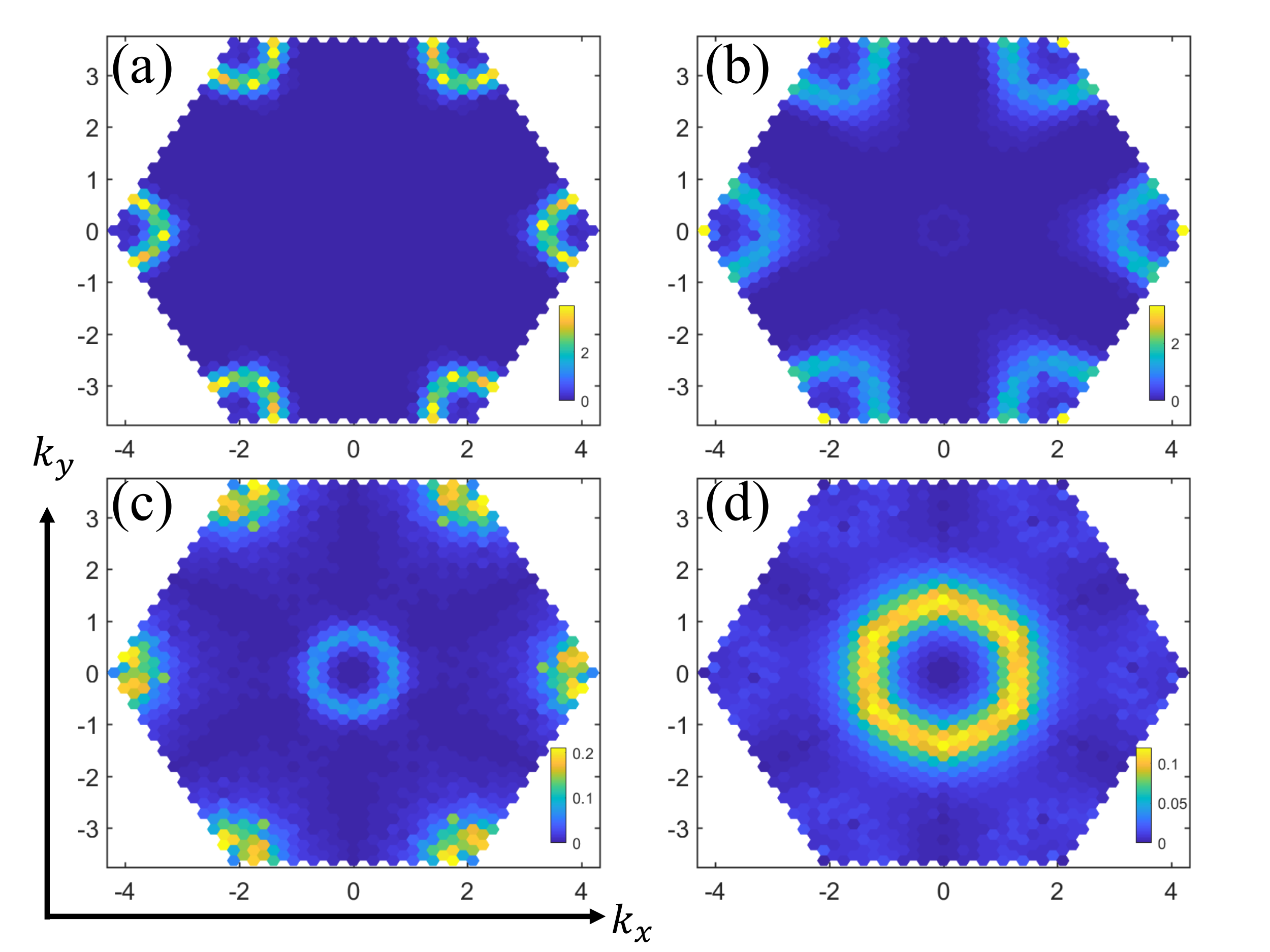}
	\caption{\label{fig2}{The tomographic slices of DSF at DQCP calculated by QMC-SAC}. The $\omega$ are fixed to (a-d) 0.08, 0.10, 0.20, and 0.40. The system size is $L=36$ with inverse temperature $\beta=600$. Because the DSF only can have a non-zero value at the reciprocal lattice site in PBC, the color in each tiny hexagon marks the value of DSF at $\textbf{k}$ of the center of the hexagon.}
\end{figure}

The lower branch (named K-branch) is largely changed and no longer flat, but the ``selection rule" still holds. Another gapless linear mode appears at K points where the order parameter of the VBS phase stays. Most strikingly, it has a different speed from $\Gamma$-branch, and can be clearly distinguished even with the naked eye. The dispersion of $\Gamma$-branch is faster than K-branch. This implies that the low-energy effective theory cannot be invariant under Lorentz transformations with one velocity. To obtain more details at the DQCP, we calculate the DSF over the entire first Brillouin zone. Then, the tomographic slices with fixed $\omega$ can be obtained and used for scanning the dynamics at DQCP.

The K-branch has a higher intensity at low $\omega$ in comparison with $\Gamma$-branch. In Fig.\ref{fig2}(a), it is nearly isotropic around the K point. However, when the energy scale increases, it becomes anisotropic (Fig.\ref{fig2}(b)) and the U(1) symmetry is broken down to Z$_3$.  Actually, in previous work \cite{dqcp_zhang,z3}, the histogram of the order parameter at DQCP does not exhibit the perfect emergent U(1) symmetry expected in theory. According to the NCCP$^1$ theory, the anisotropy may result from the three-fold monopoles which are highly close to marginal \cite{z3,z3_2}. From the dynamic spectra, we think that the possible emergent U(1) symmetry of VBS order parameter stays at a very low energy scale $\sim0.09V$, so the elimination of the anisotropy requires an extremely low temperature and large system size. At high $\omega$ in Fig.\ref{fig2} (c-d), the DSF around the K point becomes messy or weaker, and no information of the K-branch can be further extracted. In general, as $\omega$ increases, the $\Gamma$-branch becomes more clear while the K-branch is weakened. The $\Gamma$-branch is more isotropic.

\begin{figure}[t]
	\centering
	\includegraphics[width=0.48\textwidth]{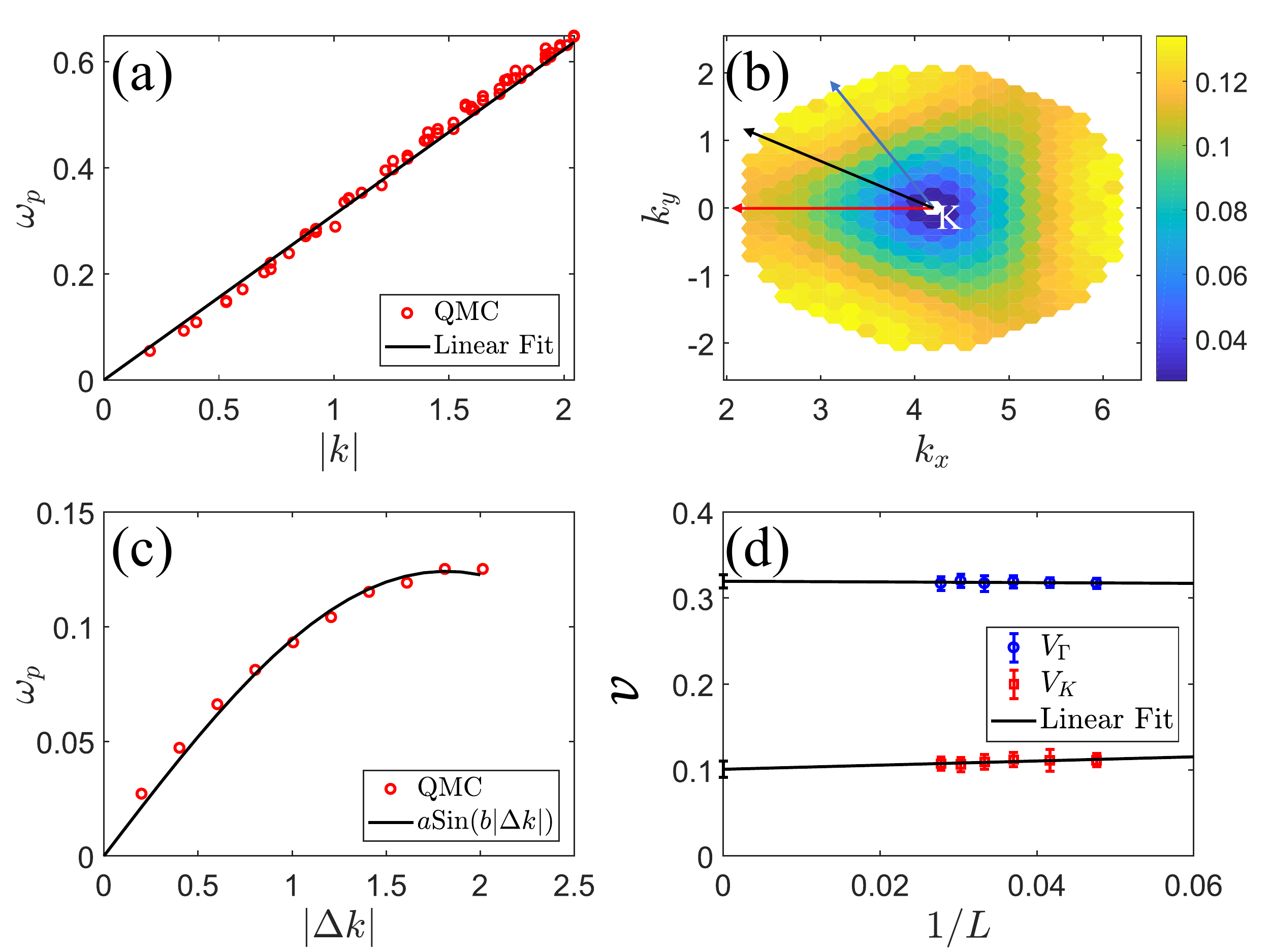}
	\caption{\label{fig3} The position of the peak $\omega_p(\textbf{k})$ of {(a)} the $\Gamma$-branch, {(b)} the K-branch, and {(c)} along angular bisector direction (black arrow in (b)). The speed of K-branch is obtained by fitting the data along the angular bisector direction between the $\Gamma$-K direction (blue arrow) and M-K direction (red arrow). The system size is $L=36$ with inverse temperature $\beta=600$. {(d)} The finite-size scaling analysis of two speeds is performed with linear fitting the data at {$L=21,24,27,30,33,36$} and $\beta=\frac{100L}{6V}$. The error bars on the $v$-axis show the fitted values for the two velocities in the thermodynamic limit.}
\end{figure}

{\underline{Two velocities}}---The velocities of different modes can be extracted from the peak position of DSF $\omega_p(\textbf{k})$ at different momentum $\textbf{k}$. Here, to avoid introducing an additional error of unnecessary fitting, we take $\omega_p(\textbf{k})$ at which the DSF reaches its local maximum at fixed $\textbf{k}$. Figure \ref{fig3} (a) plots the relation between $\omega_p(\textbf{k})$ and the magnitude of $\textbf{k}$ for the $\Gamma$-branch. We can find that the $\omega_p(\textbf{k})$ depends little on the angle of $\textbf{k}$, which firmly demonstrates the isotropy of the $\Gamma$-branch. Meanwhile, the numerical data of $\omega_p(\textbf{k})$ are in good agreement with the linear fitting result, where we obtain the constant speed of the linear dispersion at $\Gamma$ point, $V_{\Gamma}=0.317(8)$ ($L=36$).

The dispersion of the K-branch is more complicated. In Fig.\ref{fig3}(b), the DSF around the K point presents a clear feature of anisotropy. {The fastest velocity is along the M-K direction (blue arrow), and the slowest one is along $\Gamma$-K direction (red arrow).} However, both directions have only a few data points, e.g., five for $L=36$. In contrast, there are more data points along the angular bisector direction (black arrow), e.g. ten at $L=36$. In order to obtain the speed of K-branch $V_K$ with high accuracy, we fit the data along the angular bisector direction. Different from the $\Gamma$-branch, the higher energy part of the K-branch exhibits a large deviation from the linearity.  Actually, as shown in Fig.\ref{fig3}(c), we find that the sine function $\omega_{p}(\Delta\textbf{k}) = a\sin{b|\Delta\textbf{k}|}$ ($\Delta\textbf{k}=\textbf{k}-(4\pi/3,0)$) exhibits a better fit than the linear fit. Then, the speed of K-branch can be calculated by $V_K=|ab|$. Furthermore, we want to emphasize that the anisotropy does not bring serious deviation. For example, the speeds in fast, slow, and angular bisector directions at $L=36$ are 0.114(15), 0.103(21), and 0.107(8), respectively.

The large difference between the two speeds of the $\Gamma$-branch and the K-branch suggests that they originate from different physical mechanisms. First, however, we have to perform a finite-size scaling analysis to exclude possible renormalized prefactors. In Fig.\ref{fig3} (d), $V_{\Gamma}$ and $V_K$ at different system size are shown. We can find that the finite size effects are very weak, and the numerical data matches well with the linear fitting results. Finally, in the thermodynamic limit, the speed of $\Gamma$-branch is 0.319(8), and K-branch is 0.101(9). As mentioned before, it means that there are two types of gapless quasi-particle excitations with completely different speeds, therefore the Lorentz symmetry can not emerge.

\begin{figure}[t]
	\centering
	\includegraphics[width=0.48\textwidth]{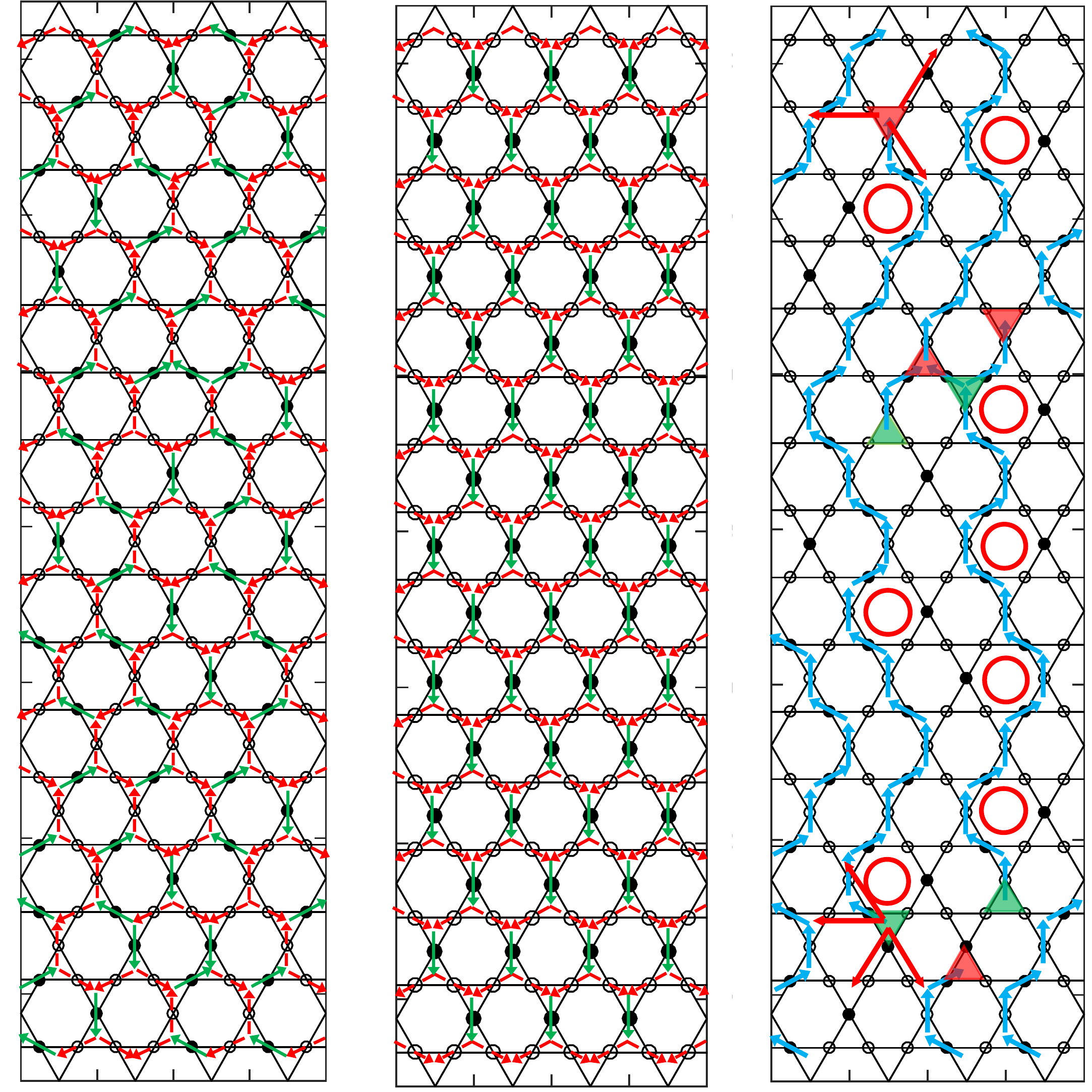}
	\caption{\label{fig4}{ A snapshot of a typical configuration during a Monte Carlo run is shown in the left panel.} The middle panel presents the reference vacuum state. The red (green) arrows mark the ``electric field" with a value of -1/3 (2/3). After subtracting the vacuum state, the open string (``electric field line") linked with the spinon (``electric charge") is shown in the right panel. The red (green) triangle labels the spinon taking negative (positive) charge one. The red arrows in the right panel show the possible directions of the spinon motion. The blue arrows mark the ``electric field" with value one. The red cycles label the hexagon where the ring exchange term can deform the string.}
\end{figure}

{\underline{Phenomenological Analysis}}---At DQCP, it is worth introducing the lattice gauge field mapping to understand the physical mechanism of topological excitations, especially the spinon (gauge charge) and the string (field line) \cite{kagome_zhang,dqcp_zhang}. As demonstrated in the left panel of Fig.\ref{fig4}, the hard-core boson can be mapped into the ``electric field" defined at the bisector of corner-shared triangle via the relation $E_{ll'} = n_{i} - 1/3$, where $l$ is located at the center of the triangle and labels the site of the dual honeycomb lattice. The gauge charges sit on the sites of dual lattice and can be calculated by summing over all field lines around one triangle via the ``Gauss law". Meanwhile, the configurations satisfying the triangle rule are the pure gauge field, and its dynamics are mainly controlled by the ring exchange term which acts as the ``magnetic field" \cite{ice_Balents,ice_sac,dqcp_zhang}. At the DQCP, the triangle rule is broken, and the spinons emerge. To make them better illustrated, we set the stripe state (Fig.\ref{fig4} middle panel) as the reference vacuum state (uniform ``electric field"). Then, after subtracting the vacuum from the snapshot of QMC configuration (Fig.\ref{fig4} left panel), the spinons connected with the quantum string can be clearly observed (Fig.\ref{fig4} right panel).

In the language of lattice gauge field, the nearest-neighbor hopping of bosons on the Kagome lattice can be transformed into the next-nearest-neighbor hopping of the spinons in the dual honeycomb lattice (red arrows in Fig.\ref{fig4} right panel). Because the honeycomb lattice is bipartite, the up-triangle and down-triangle reside in different sublattices, and they corresponds to different type of spinons. Thus, the low energy physics can be approximately understood as the dynamic U(1) gauge field coupled to four types of spinons: 
\begin{center}
	\begin{tabular}{|c|c|c|c|c|}
		\hline
		Type: & ${\tikz{\draw (0,0) -- (0.1,0.1732) -- (0.2,0) -- (0,0);}}$ 
		& ${\tikz{\fill (0,0) circle (0.5mm);\fill (0.1,0.1732) circle (0.5mm) ;\draw (0,0) -- (0.1,0.1732) -- (0.2,0) -- (0,0);}}$, 
		${\tikz{\fill (0.2,0) circle (0.5mm);\fill (0.1,0.1732) circle (0.5mm) ;\draw (0,0) -- (0.1,0.1732) -- (0.2,0) -- (0,0);}}$,
		${\tikz{\fill (0.2,0) circle (0.5mm);\fill (0.0,0.0) circle (0.5mm) ;\draw (0,0) -- (0.1,0.1732) -- (0.2,0) -- (0,0);}}$ 
		& ${\tikz{\fill (0.1,0) circle (0.5mm);\fill (0,0.1732) circle (0.5mm) ;\draw (0,0.1732) -- (0.2,0.1732) -- (0.1,0) -- (0,0.1732);}}$,
		${\tikz{\fill (0.1,0) circle (0.5mm);\fill (0.2,0.1732) circle (0.5mm) ;\draw (0,0.1732) -- (0.2,0.1732) -- (0.1,0) -- (0,0.1732);}}$,
		${\tikz{\fill (0.2,0.1732) circle (0.5mm);\fill (0,0.1732) circle (0.5mm) ;\draw (0,0.1732) -- (0.2,0.1732) -- (0.1,0) -- (0,0.1732);}}$ 
		& ${\tikz{\draw (0,0.1732) -- (0.2,0.1732) -- (0.1,0) -- (0,0.1732);}}$\\
		\hline
		Charge: & +1 & -1 & +1 & -1 \\
		\hline
		Field: & $p_1$ & $h_1$ & $p_2$ & $h_2$\\
		\hline
	\end{tabular}
\end{center}
As demonstrated in right panel of Fig.\ref{fig4}, the possible hopping directions of spinon are not fixed, but we can set an average hopping amplitude $t'=7t/12$ for all the next-nearest-neighbor directions. Then with assumption of free spinons, the energy dispersion can be obtained $E(\textbf{k})=-2 t'\left(2 \cos \left(\frac{\sqrt{3} k_x}{2}\right) \cos \left(\frac{k_y}{2}\right)+\cos (k_y)\right)\approx -t(\frac72-\frac78k^2)$.  Similar to the Mott-SF phase transition \cite{Balents2005ptps}, we can define the pair operator $\Psi_\alpha=(p_\alpha+h_\alpha^\dagger)/\sqrt{2}$ and $\Xi_\alpha=(p_\alpha-h_\alpha^\dagger)/\sqrt{2}$. Because the DQPT happens at exact 1/3 filling, the densities of spinons $p_\alpha$ and $h_\alpha$ are the same. Then after integrating out $\Xi_\alpha$, the effective Lagrangian becomes ``relativistic" and the velocity of $\Gamma$-branch can be estimated $V_k\sim\sqrt{\frac78t_c}=0.3375$ which is close to the numerical result. Therefore, we conclude that the gapless mode of $\Gamma$-branch may be caused by the deconfinement of two types of ``spinon pair" (The spectra of off-diagonal structure factor and spinon density correlation also support that).

On the other hand, the K-branch is the deformation of the flat band, so it should be relevant to the effective ring exchange interaction. In previous work \cite{kagome_zhang}, the quantum string can be well described by the spin half XY chain at half filling. The ring exchange term is equivalent to the XY spin exchange interaction with the same strength $t_{e} = -\frac{12t^{3}}{V^{2}}$. It is well known that the excitations of spin half XY chain are kink anti-kink or the free Jordan-Wigner fermions. Then near the Fermi point, the velocity of dispersion is $2t_e$. Because the primitive vector of the string is $\sqrt{3}/2$ times the lattice vector, the corresponding velocity should be rescaled to $4t_e/\sqrt{3}=0.0612$ which is apparently small. Unlike Ref.\cite{kagome_zhang}, the quantum strings at DQCP are open strings with spinons attached to the ends, so the velocity should be strongly affected by the complex interplay between spinons and quantum strings.

\underline{Conclusion and Discussion}---From the dynamic spectra of DQPT on the Kagome lattice, two linear dispersions are observed. The dispersion of spinons is fast and contributes to the linear dispersion of $\Gamma$-branch, and the slow linear mode at K point may result from the internal excitation of the open quantum strings. Due to the large difference between the two speeds, the Lorentz symmetry cannot emerge at DQCP.

The DQPT on the Kagome lattice is continuous, so it is not likely to be the non-unitary CFT which usually results in the weakly first order. One possibility of the CFT is that the marginal terms may make the velocities of the two modes to be different. Alternatively, similar to the two velocities caused by the spin-charge separation in the one-dimensional fermionic model \cite{spin_charge}, the decoupled CFT method may be also suitable for analyzing the DQPT. Recently, there is a novel understanding of DQPT via the generalized higher-form symmetries \cite{higher-form} which may provide more novel understanding into these issues.

\section*{acknowledgments}
We thank Yin-Chen He, Hong-Hao Tu, Meng Cheng, Chang-Le Liu, Zheng Yan, Yan-Cheng Wang, Jin Zhang, and Zheng Zhou very much for their valuable discussions and suggestions. X.-F. Z. acknowledges funding from the National Science Foundation of China under Grants  No. 12274046, No. 11874094, No.12147102, and No.12347101, Chongqing Natural Science Foundation under Grants No. CSTB2022NSCQ-JQX0018, Fundamental Research Funds for the Central Universities Grant No. 2021CDJZYJH-003, and Xiaomi Foundation / Xiaomi Young Talents Program. Zijian Xiong acknowledges funding from the International Postdoctoral Exchange Fellowship Program 2022 by the Office of China Postdoctoral Council (Grant No. PC2022072) and the National Natural Science Foundation of China (Grant No. 12147172). Yining Xu is supported by the Natural Science Foundation of Chongqing(Grant No.cstc2021jcyj-msxmX0428) , and by the Science and Technology Research Program of Chongqing Municipal Education Commission (Grant No. KJQN202100514)

\section*{Appendix}
\section{Finite Size Scaling of the Energy Gap}
\begin{figure}[t]
	\centering
	\includegraphics[width=0.5\textwidth]{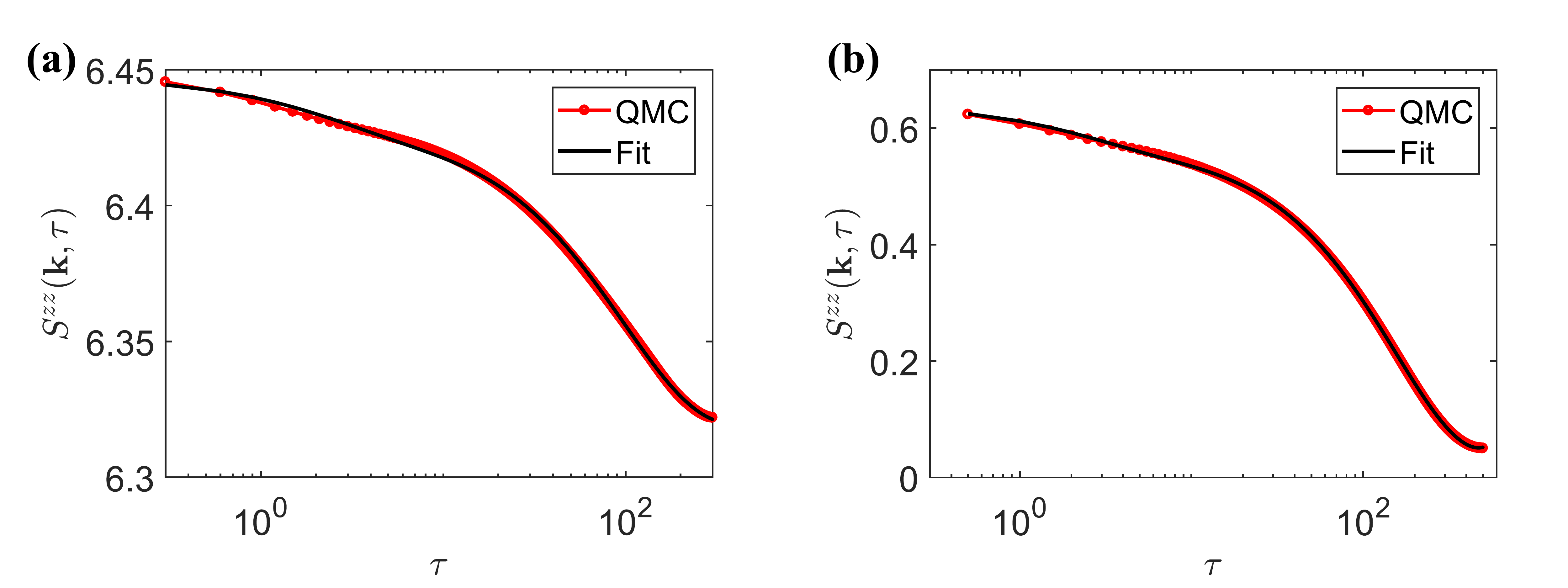}
	\caption{The imaginary correlation function $S^{zz}(\mathbf{k},\tau)$ of (a) VBS phase ($t/V=0.08$) and (b) DQCP ($t/V=0.1302$) with $L=36$. Due to the periodical boundary condition in the imaginary time, the region of $\tau$ is $[0 $ $\beta/2]$. The red dot lines are the numerical result of QMC and the black lines are fitting curves.\label{figs1}}
\end{figure}

The energy gaps of different phases can be directly calculated from the imaginary correlation function in the momentum space $S^{zz}(\mathbf{k},\tau)=\langle e^{\tau H}S^z(\mathbf{k})^{\dagger}e^{-\tau H}S^z(\mathbf{k})\rangle$. Assuming the eigenstates are $|i\rangle$, then $S^{zz}(\mathbf{k},\tau)$ can be written as
\begin{eqnarray}
\nonumber
&&S^{zz}(\mathbf{k},\tau)=\frac 1 Z\sum_i\langle i|e^{-(\beta-\tau)H}S^z(\mathbf{k})^{\dagger}e^{-\tau H}S^z(\mathbf{k})|i\rangle\\
\nonumber
&=&\frac 1 Z\sum_{i,j}e^{-(\beta-\tau)E_i}e^{-\tau E_j}|\langle i|S^z(\mathbf{k})|j\rangle|^2 \\
\nonumber
&=&\frac 1 Z(\sum_{i}e^{-\beta E_i}|\langle i|S^z(\mathbf{k})|i\rangle|^2 \\
\nonumber
&&+\sum_{i<j}(e^{-(\beta-\tau)E_i}e^{-\tau E_j}+e^{-(\beta-\tau)E_j}e^{-\tau E_i})|\langle i|S^z(\mathbf{k})|j\rangle|^2) \\
\nonumber
&=&S^{zz}_0(\mathbf{k})+\frac 1 Z\sum_{i<j}e^{-\beta E_i}|\langle i|S^z(\mathbf{k})|j\rangle|^2(e^{-\tau\Delta E_{ij}}+e^{-(\beta-\tau)\Delta E_{ij}})\\
\nonumber
&\approx&S^{zz}_0(\mathbf{k})+\frac {e^{-\beta E_0}} Z \sum_{j\ne 0}|\langle 0|S^z(\mathbf{k})|j\rangle|^2(e^{-\tau\Delta E_{0j}}+e^{-(\beta-\tau)\Delta E_{0j}}),
\end{eqnarray}
where $S^{zz}_0(\mathbf{k})=\frac {\sum_{i}e^{-\beta E_i}|\langle i|S^z(\mathbf{k})|i\rangle|^2} Z$ is residual value at large $\tau$ and $\Delta E_{ij}$ is the energy difference. Thus, to get the lowest energy gap, we fit $S^{zz}(\mathbf{k},\tau)$ with ansatz $f(t)=c_0+a_1(e^{-\tau \Delta_1}+c_1e^{\tau \Delta_1})+a_2e^{-\tau \Delta_2}$. As demonstrated in Fig.\ref{figs1}, the numerical results match very well with the fitting curves. The fast decaying at small $\tau$ is related to term $a_2e^{-\tau \Delta_2}$ with higher energy gap $\Delta_2$. The lowest energy gap corresponds to the slow decaying at large $\tau$ and is related to term $a_1(e^{-\tau \Delta_1}+c_1e^{\tau \Delta_1})$. Here, introducing term $c_1e^{\tau \Delta_1}$ is due to the consideration of the effect of $e^{-(\beta-\tau)\Delta E_{0j}}$ around $\tau=\beta/2$. 

\begin{figure}[h]
	\centering
	\includegraphics[width=0.5\textwidth]{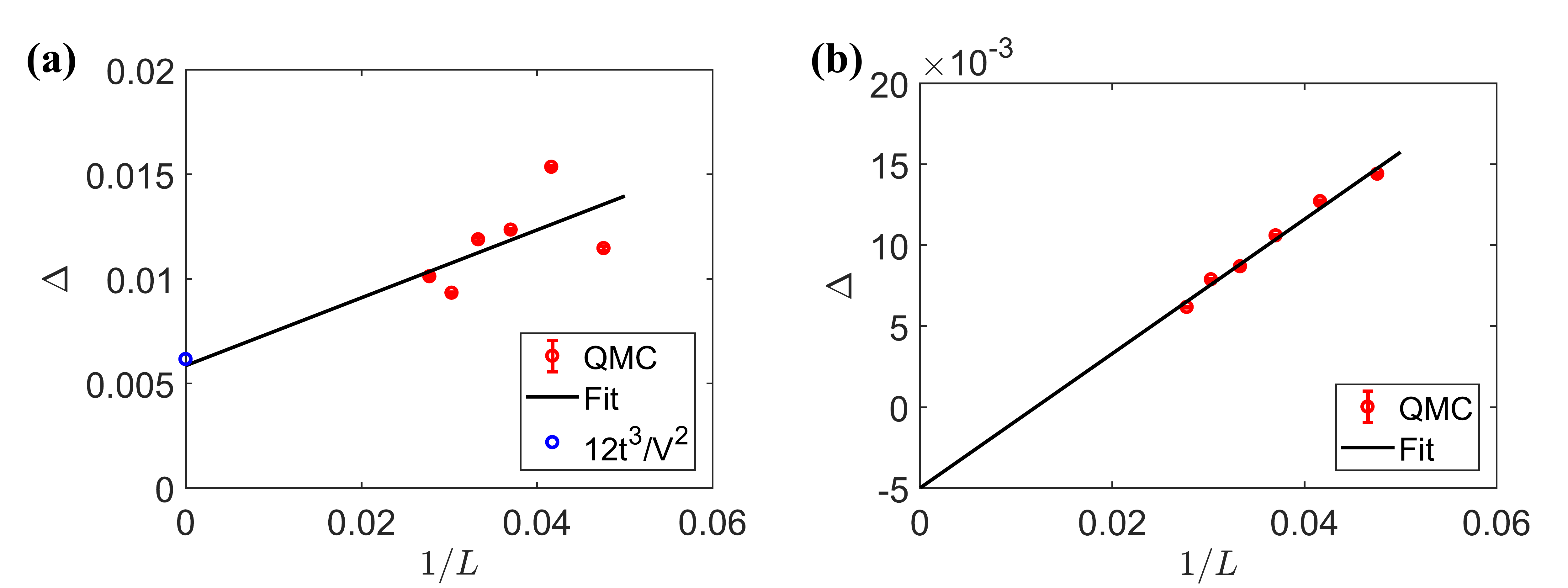}
	\caption{The lowest energy gap at K point of (a) VBS phase and (b) DQCP. The red dot lines are the numerical result of QMC and the black lines are linear fitting results.\label{figs2}}
\end{figure}

In the finite system, there is a small energy gap even in the gapless system. Thus, we make use of the finite-size scaling to check the energy gap in the thermodynamic limit. In the VBS phase (Fig.\ref{figs2}(a)), the energy gaps at different sizes are larger than $12t^3/V^2$. After linear fitting, the energy gap is still positively large. Here, we have to mention that the results of VBS present a strong deviation from the fitting results, and it may result from the finite size effects. In comparison, at DQCP (Fig.\ref{figs2}(b)), the gap energies at different system sizes exhibit good linear behavior. The negative value at the thermodynamic limit indicates the close of the energy gap. 

\section{Dynamical Spectra}
\begin{figure}[t]
	\centering
	\includegraphics[width=0.5\textwidth]{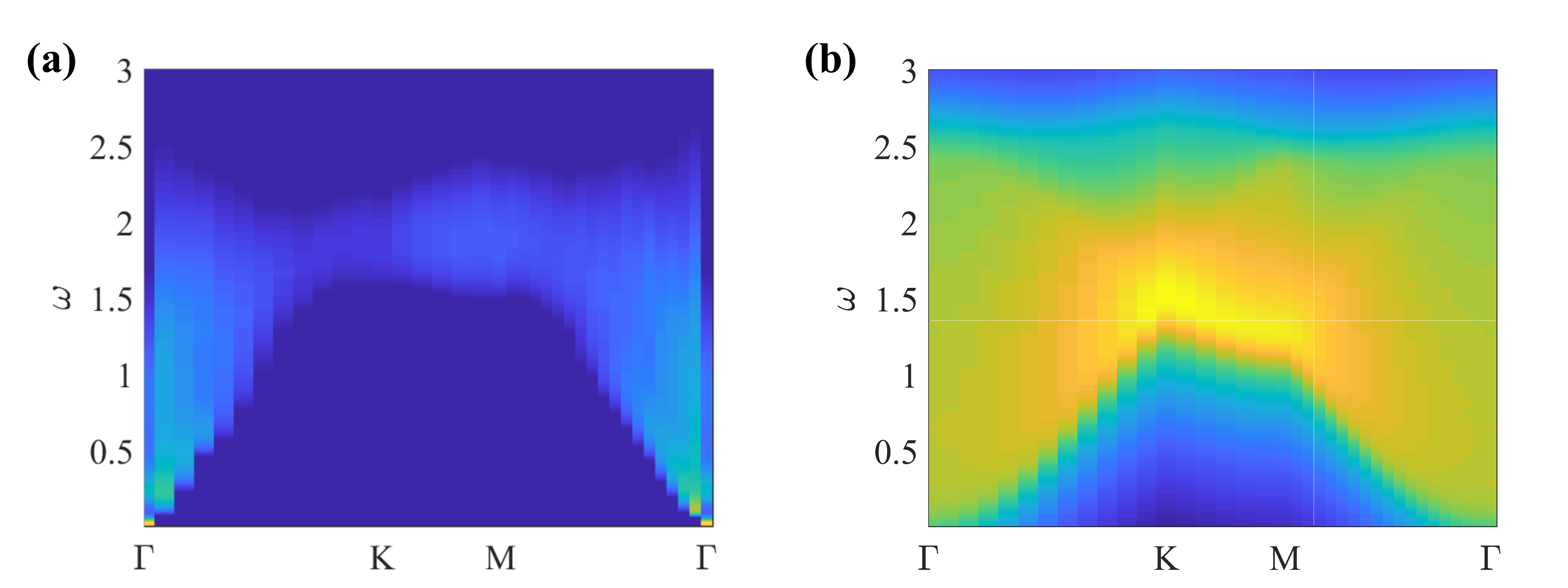}
	\caption{\label{figs3} The dynamic spectrum of xy-component $S^{+-}(\mathbf{k},\omega)$ at DQCP with $L=36$, calculated by the (a) QMC and (b) tight-binding free spinon model. The intensity is plotted in a logarithmic scale.}
\end{figure}

The off-diagonal (or xy-component) dynamical structure factor can be written as $$S^{+-}(\textbf{k},\omega)=\frac{1}{2\pi L^2}\sum_{ij}\int_{-\infty}^{+\infty}\mbox{d}t\,e^{i\mathbf{k}\cdot(\mathbf{r}_{i}-\mathbf{r}_{j})-i\omega t}\langle S_{i}^{+}(0)S_{j}^{-}(t)\rangle.$$ As shown in Fig.\ref{figs3}(a), the features of the spectrum are quite similar to the BFG model in Kagome lattice \cite{BFG_Wessel} and the quantum spin ice in pyrochlore lattice \cite{CJHuang2018prl}. In both cases, the off-diagonal spectrum can be understood with the help of the tight-binding free spinon model. In the language of lattice gauge field theory, the spin operator $S_{i}^{+}$ is two body operator of spinons, so the off-diagonal dynamical structure factor can be written as $S^{+-}_{\alpha\beta}(\textbf{k},\omega)=\frac{1}{N}\sum_q\delta(\omega-\epsilon_q-\epsilon_{-q-k})\times e^{-i(k+2q)\cdot (r_{\alpha}-r_{\beta})}$ where $r_{\alpha}$ labels the relative direction between dual sites. After substituting the delta function with the Lorentzian function and assuming the spinons can freely move on the dual lattice via strong next-nearest-neighbor hopping, we analytically obtain the off-diagonal spectrum shown in Fig.\ref{figs3}(b). The envelope lines approximately agree with the numerical results, and the mismatch of intensity may result from the exotic distribution of spinons.

\begin{figure}[h]
	\centering
	\includegraphics[width=0.45\textwidth]{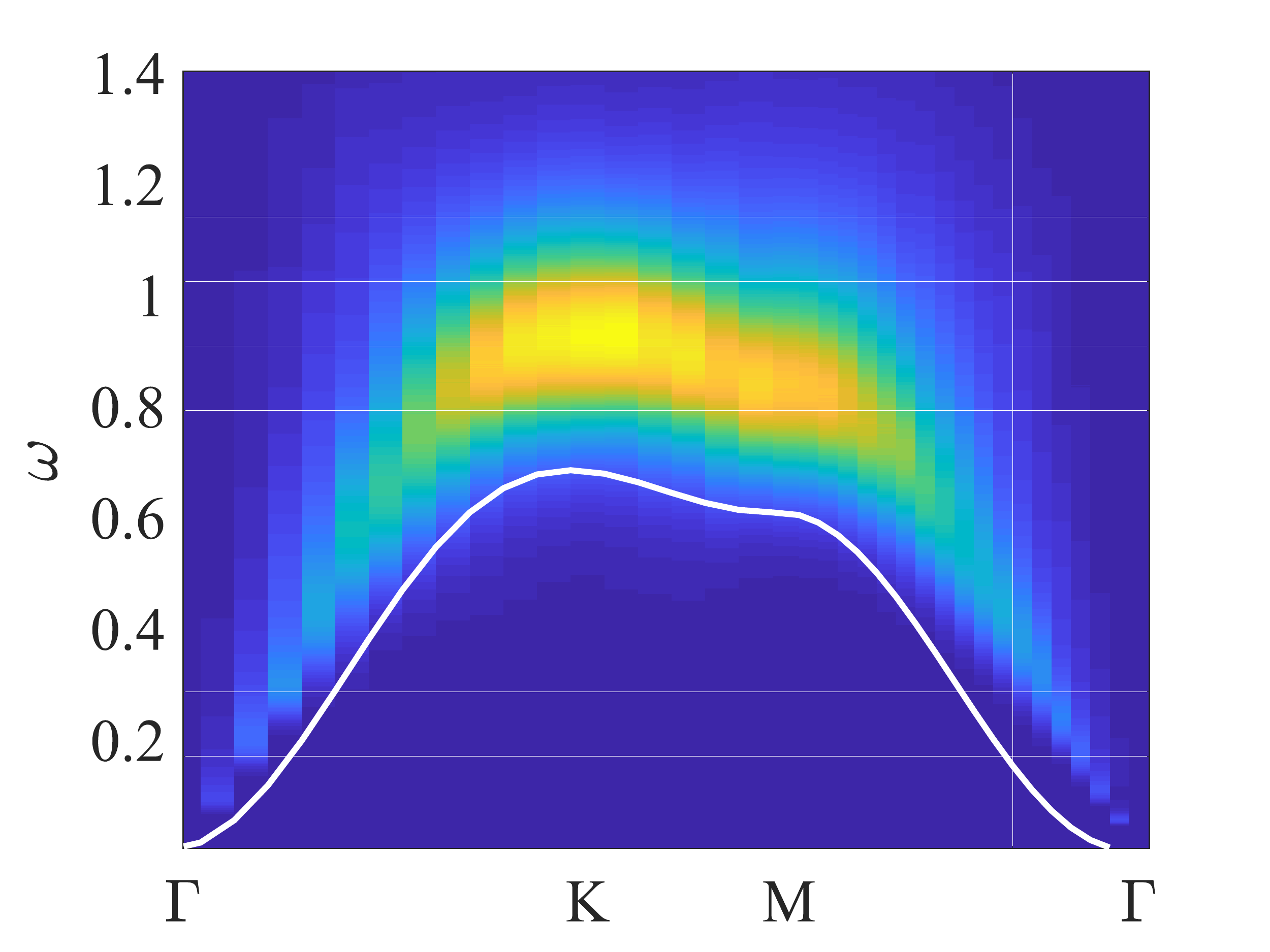}
	\caption{\label{figs4} The dynamic spectrum of spinon at DQCP with $L=36$. The white line is the dispersion of spinon calculated via the tight-binding free spinon model.}
\end{figure}

On the other hand, we also calculate the spectrum of spinon density correlation function $\langle S^{\triangle}_l(\tau)S^{\triangle}_{l'}(0)\rangle$. Same as the definition in our previous paper \cite{dqcp_zhang}, $S^{\triangle}_l=\sum_{i\in \triangle_l}S^z_i$ is the density of spinon or gauge charge at the dual lattice. In Fig.\ref{figs4}, we can find there is only one branch staying at a higher energy scale. Compared with the zz-component spectrum, we can find the disappearance of the lower branch corresponds to the dynamics of the string. It means the spectrum of spinon density correlation function can only reflect spinons' dynamics.

\bibliographystyle{apsrev4-1}
\bibliography{ref}
\end{document}